\documentclass{article}
\usepackage{amssymb}

\usepackage{sw20ssur}

\input{tcilatex}

\begin{document}

\author{Azhar Iqbal \\
{\small Department of Mathematics, University of Hull, Hull, HU6 7RX, UK}}
\title{Playing games with EPR-type experiments}
\date{August 20, 2005}
\maketitle

\begin{abstract}
An approach towards quantum games is proposed that uses the unusual
probabilities involved in EPR-type experiments directly in two-player games.
\end{abstract}


\section{Introduction}

The emerging field of quantum games \cite
{Vaidman,Meyer,Eisert,Eisert1,Benjamin,Benjamin1,Marinatto,Enk,Piotrowski,Piotrowski1,Du,Du1,Du2,Piotrowski2,Flitney1,Flitney2,Cheon,Shimamura,Johnson,Flitney,Iqbal,Orlin,Aerts,grib}%
\ has attracted increasing attention during recent years. Some authors \cite
{Benjamin1,Enk,Aerts,Cheon} have pointed out that, in certain cases, it is
also possible to construct classical description of quantum games, resulting
in the presently continuing debate \cite{Iqbal1,Aerts,Cheon}\ about their
true quantum content and character.

Game theory is based on, and extensively uses, the theory of probability 
\cite{Chung}. The peculiar probabilities arising in the EPR-type experiments 
\cite{EPR,CHSH,CHSH1,CHSH2} are considered the basis of arguments for
quantum non-locality \cite{Peres}. Sometimes, these probabilities are also
presented as the most convincing demonstration of how quantum and classical
mechanics differ from each other. The motivation behind EPR-type experiments
is the EPR paradox \cite{EPR}. In an attempt to resolve the EPR paradox M%
\"{u}ckenheim \cite{Muckenheim}\ in 1982 made use of negative probability
functions. Moreover, recent years have witnessed explicit proofs \cite{Han
et al,Cereceda,Rothman} showing how certain probability measures involved in
some local hidden variable (LHV) models of the EPR paradox attain negative
values. Quoting M\"{u}ckenheim \cite{Muckenheim1} ``\textit{Kolmogorov's
axiom may hold or not; the probability for the existence of negative
probabilities is not negative.}'' Also, Feynman \cite{Fenyman,Feynman1} once
cleverly anticipated \cite{Cereceda,Rothman} that `` The \emph{only}
difference between the classical and quantum cases is that in the former we
assume that the probabilities are positive-definite.'' The physical meaning
of the negative probability measures is indeed far from obvious and this
particular attempt to resolve the EPR paradox is taken \cite{Muckenheim2} ``%
\textit{as unattractive as all others.''}

In present paper we adapt a positive attitude towards negative probabilities
by observing that, in spite of their unattractiveness, they seem to have
undisputed value in terms of providing an indication to what can be taken as
the true quantum character in certain quantum mechanical experiments.
Secondly, the negative probabilities, though labelled as unattractive, have
a potential to provide an alternative, and perhaps much shorter, route to
construction of simple examples of quantum games. These constructions
designed for the physical implementation, and realization, of quantum games
will, of course, have the EPR-type experiments as their underlying physical
structure.

The approach towards quantum games, developed in this paper, can be divided
into two steps. In the first step, elementary probability theory is used to
analyze a hypothetical physical implementation of a bi-matrix game, that is
motivated by, and has close similarities with, the experimental set-up of
the EPR-type experiments. Looking from a distance, this analysis can also be
taken as a procedure that re-defines the classical game in a way that
justifies and opens the way towards the next step in the present approach.
In the second step the peculiar probabilities, coming out of the EPR-type
experiments are introduced to see their resulting effects on players' payoff
functions and solutions of the game.

Apart from a being a short-cut towards demonstration and construction of
simple quantum games, to us this approach seems to be in demand presently
both in game theory and economics \cite{Piotrowski2}. Recent years have
witnessed serious efforts to entertain the methods and notation of quantum
mechanics in these domains. In our view, in spite of these developments, it
remains a fact that in these domains the concepts of wavefunction, ket bra
notation, and Hermitian operators are still associated with an alien
identity; with quantum mechanics believed to be their \emph{only} right
place. Present paper tries to fill in this gap by looking at how quantum
games can also be understood by using the peculiar probabilities that appear
in certain well-known quantum mechanical experiments, without recourse to
the mathematical tools of quantum mechanics. In other words, we try to show
how the unusual probabilities in EPR paradox have a potential to leave their
mark on game theory.

The rest of this paper is organized as follow. Section (\ref{Physical
implementation of a bi-matrix game}) compares playing bi-matrix game with
the setting of EPR-type experiments to motivate a four-coin physical
implementation of a bi-matrix game. Section (\ref{Games with four coins})
develops such a hypothetical physical implementation. Section (\ref{Games
with perfectly correlated}) builds up on the construction of the previous
section to look at how the game is affected when, instead of four coins, two
correlated particles are used to play the game.

\section{\label{Physical implementation of a bi-matrix game}Physical
implementation of a bi-matrix game}

We consider a two-player two-strategy non-cooperative game that is given as
a bi-matrix. Two players Alice and Bob are not allowed to communicate and
each player has to go for one of the two available strategies. A usual
physical implementation of this game consists of giving two coins to Alice
and Bob, each receiving one. Both receive the coin in head state. Each
player plays his/her strategy that consists of flipping/not flopping the
coin. Afterwards, the players return their coins to a referee. The referee
observes the coins and rewards the players according to the strategies they
have played and the game under consideration.

Consider now the well-known setting of EPR-type experiments. Once again,
Alice and Bob are spatially separated and are unable to communicate with
each other. Both receive one half of a pair of particles that originate from
a common source. In an individual run both choose one from the two options
(strategies) available to him/her. The strategies are usually two directions
in space along which measurements can be made. Each measurement generates $%
+1 $ or $-1$ as the outcome, which can be associated to head and tail states
of a coin, respectively. Experimental results are recorded for a large
number of individual runs of the experiment.

Apparent similarities between the two-coin physical implementation of a
bi-matrix game and EPR-type experiments can immediately be noticed. The
similarities hint to use EPR-type experiments to play bi-matrix games.
However, before moving further along that direction, following observations
are made:

\begin{enumerate}
\item  In two-coin implementation of a bi-matrix game, a player knows the
head or tail state of his/her coin after he/she has played his/her strategy.

\item  In EPR-type experiment when a player decides his/her strategy, as one
of two available directions along which a measurement is to be made, he/she
does not know whether the measurement is going to result in $+1$ or $-1$,
until the measurement has actually been made.
\end{enumerate}

It shows that a two-coin physical implementation of a bi-matrix game is not
a right analogy to EPR-type experiments. In an individual run of the
EPR-type experiment each player has to chose from one of two available
directions. After the player makes a choice between the two directions, the
measurement generates $+1$ or $-1$ as outcome. It motivates a four-coin
implementation of the game.

\section{\label{Games with four coins}Bi-matrix games with four coins}

A bi-matrix game can also be played using four coins instead of the two
described above. A procedure that physically implements the game with four
coins is as follows. Its motivation comes from the EPR-type experiments and
it serves to make possible, in the next step, a smoother transition towards
a situation when the \emph{same} game is played using those experiments.

The players Alice and Bob are given two coins each. It is not known, and it
does not matter, whether the given coins are in head or tail states. Before
the game starts the referee announces that a player's strategy is to choose
one out of the two coins in his/her possession. After playing their
strategies, the players give the two chosen coins, representing their
strategies, to the referee. The referee tosses both the coins together and
records the outcomes. The tossing of the coins is repeated a large number of
times, while each player plays his/her strategy each time he/she is asked to
choose one of the two coins in his/her possession. After a large number of
individual runs the referee rewards the players according to the strategies
they have played and the probability distributions followed by the four
coins during their tossing.

After stating the general idea, we consider in the following an example of a
symmetric bi-matrix game.

\begin{equation}
\begin{array}{c}
\text{Alice}
\end{array}
\begin{array}{c}
X_{1} \\ 
X_{2}
\end{array}
\stackrel{\stackrel{
\begin{array}{c}
\text{Bob}
\end{array}
}{
\begin{array}{cc}
\acute{X}_{1} & \acute{X}_{2}
\end{array}
}}{\left( 
\begin{array}{cc}
(K,K) & (L,M) \\ 
(M,L) & (N,N)
\end{array}
\right) }  \label{matrix}
\end{equation}
where $X_{1}$, $X_{2}$, $\acute{X}_{1}$ and $\acute{X}_{2}$ are the players'
strategies. Entries in braces are Alice's and Bob's payoffs, respectively.

We want to physically implement the game (\ref{matrix}) using repeated
tossing of four coins which follows the following probability distribution

\begin{equation}
\begin{array}{c}
\text{Alice}
\end{array}
\stackrel{
\begin{array}{c}
\text{Bob}
\end{array}
}{
\begin{array}{cc}
\begin{array}{c}
S_{1}
\end{array}
\begin{array}{c}
H \\ 
T
\end{array}
\stackrel{\stackrel{
\begin{array}{c}
\acute{S}_{1}
\end{array}
}{
\begin{array}{cc}
H & T
\end{array}
}}{\left( 
\begin{array}{cc}
p_{1} & p_{2} \\ 
p_{3} & p_{3}
\end{array}
\right) } & \stackrel{\stackrel{
\begin{array}{c}
\acute{S}_{2}
\end{array}
}{
\begin{array}{cc}
H & T
\end{array}
}}{\left( 
\begin{array}{cc}
p_{5} & p_{6} \\ 
p_{7} & p_{8}
\end{array}
\right) } \\ 
\begin{array}{c}
S_{2}
\end{array}
\begin{array}{c}
H \\ 
T
\end{array}
\left( 
\begin{array}{cc}
p_{9} & p_{10} \\ 
p_{11} & p_{12}
\end{array}
\right) & \left( 
\begin{array}{cc}
p_{13} & p_{14} \\ 
p_{15} & p_{16}
\end{array}
\right)
\end{array}
}  \label{4 Coin Stats}
\end{equation}
where, for example, $S_{1}$ is Alice's pure strategy to ``always select the
coin $1$'' etc. The pure strategies $S_{2},\acute{S}_{1}$ and $\acute{S}_{2}$
can similarly be interpreted. Also, $H\thicksim $Head and $T\thicksim $Tail
and, for obvious reasons, we have

\begin{equation}
\stackrel{4}{\stackunder{1}{\sum }}p_{i}=\stackrel{8}{\stackunder{5}{\sum }}%
p_{i}=\stackrel{12}{\stackunder{9}{\sum }}p_{i}=\stackrel{16}{\stackunder{13%
}{\sum }}p_{i}=1  \label{Constraints}
\end{equation}
In construction of the four-coin statistics (\ref{4 Coin Stats}) following
points should be taken into consideration.

\begin{enumerate}
\item  The statistics (\ref{4 Coin Stats}) may convey the impression that in
an individual run both players forward both of their coins to the referee
who tosses the four coins together. In fact, in an individual run the
referee tosses only two coins. The statistics (\ref{4 Coin Stats}) are
generated under the \emph{assumption} that there is randomness involved in
players' strategies to go for one or the other coin.

\item  Associated with the above impression is the fact that, in every
individual run, the statistics (\ref{4 Coin Stats}) assign head or tail
states to the two coins that have \emph{not} been tossed. So that, in each
individual run two tosses are \emph{counterfactual}.
\end{enumerate}

Because counterfactual reasoning is involved in the derivation of Bell-CHSH
inequality, some authors \cite{Fine,Fivel,Gustafson} have argued that
quantum non-locality \cite{Peres}\ (or locality) does not follow from the
violation of the Bell-CHSH inequality in EPR-type experiments. With
reference to our coin game counterfactual reasoning means that two coins,
out of four, are not tossed in each individual turn, but still these
untossed coins are assigned head or tail states in the mathematical steps
used in the derivation of Bell-CHSH inequality. This assignment is often
justified under the label of \emph{realism}. In EPR-type experiments the
measurements along two directions, for each player, do not commute with each
other i.e. joint measurements can not be made. Based on this fact some
arguments \cite{DeBaere,Malley1,Malley}\ say that the reasons for the
violation of Bell-CHSH inequality in EPR-type experiments reside \emph{only}
in non-commutative nature of the quantum measurements involved.

To play the game (\ref{matrix}) with four-coin statistics (\ref{4 Coin Stats}%
), we assume the referee has the following recipe\footnote{%
The recipe (\ref{Payoffs}) is not unique and others may be suggested. Any
recipe is justified if it is able to reproduce the game under consideration
within the description of the statistical experiment involved.} to reward
the players.

\begin{equation}
\left. 
\begin{array}{c}
P_{A}(S_{1},\acute{S}_{1})=Kp_{1}+Lp_{2}+Mp_{3}+Np_{4} \\ 
P_{A}(S_{1},\acute{S}_{2})=Kp_{5}+Lp_{6}+Mp_{7}+Np_{8} \\ 
P_{A}(S_{2},\acute{S}_{1})=Kp_{9}+Lp_{10}+Mp_{11}+Np_{12} \\ 
P_{A}(S_{2},\acute{S}_{2})=Kp_{13}+Lp_{14}+Mp_{15}+Np_{16}
\end{array}
\right\}  \label{Payoffs}
\end{equation}
Where $P_{A}(S_{1},\acute{S}_{2})$, for example, is Alice's payoff when she
plays $S_{1}$ and Bob plays $\acute{S}_{2}$. The corresponding payoff
expressions for Bob can be found by the transformation $L\leftrightarrows M$
in Eqs. (\ref{Payoffs}). The recipe, of course, makes sense if repeated
tosses are made with four coins. Because $S_{1},S_{2},\acute{S}_{1}$and $%
\acute{S}_{2}$ are taken as players' pure strategies; a mixed strategy for
Alice, for example, is convex linear combination of $S_{1}$ and $S_{2}$.

We now find constraints on four-coin statistics (\ref{4 Coin Stats}) such
that each equation in (\ref{Payoffs}) represents mixed strategy payoff for
the bi-matrix game (\ref{matrix}), that can be written in bi-linear form. To
allow this interpretation for the payoffs (\ref{Payoffs}) four probabilities 
$r,$ $s,$ $\acute{r}$ and $\acute{s}$ are required that can give a bi-linear
representation to the payoffs (\ref{Payoffs}) i.e.

\begin{equation}
\left. 
\begin{array}{c}
P_{A}(S_{1},\acute{S}_{1})=Kr\acute{r}+Lr(1-\acute{r})+M\acute{r}%
(1-r)+N(1-r)(1-\acute{r}) \\ 
P_{A}(S_{1},\acute{S}_{2})=Kr\acute{s}+Lr(1-\acute{s})+M\acute{s}%
(1-r)+N(1-r)(1-\acute{s}) \\ 
P_{A}(S_{2},\acute{S}_{1})=Ks\acute{r}+Ls(1-\acute{r})+M\acute{r}%
(1-s)+N(1-s)(1-\acute{r}) \\ 
P_{A}(S_{2},\acute{S}_{2})=Ks\acute{s}+Ls(1-\acute{s})+M\acute{s}%
(1-s)+N(1-s)(1-\acute{s})
\end{array}
\right\}  \label{bilinear payoffs}
\end{equation}
It then allows to make the association

\begin{equation}
S_{1}\thicksim r\text{, \ \ }S_{2}\thicksim s\text{, \ \ }\acute{S}%
_{1}\thicksim \acute{r}\text{, \ \ }\acute{S}_{2}\thicksim \acute{s}
\label{correspondences}
\end{equation}
where $r,s,\acute{r}$ and $\acute{s}$ are the probabilities of heads for
coins $S_{1},S_{2},\acute{S}_{1}$ and $\acute{S}_{2}$, respectively. In case
a consistent set of these four probabilities is found, each equation in (\ref
{Payoffs}) can be interpreted in terms of a mixed strategy game between the
two players. So that, the four pairs

\begin{equation}
(r,\acute{r}),\text{ \ }(r,\acute{s}),\text{ \ }(s,\acute{r}),\text{ \ }(s,%
\acute{s})  \label{probability pairs}
\end{equation}
represent the four possible situations that may result when each player has
got two strategies to choose from. For example, the strategy pair $(S_{1},%
\acute{S}_{2})$ is associated with the pair $(r,\acute{s})$ and it
corresponds to the mixed strategy game given as

\begin{equation}
\begin{array}{c}
\text{Alice}
\end{array}
\begin{array}{c}
r \\ 
(1-r)
\end{array}
\stackrel{\stackrel{
\begin{array}{c}
\text{Bob}
\end{array}
}{
\begin{array}{ccc}
\acute{s} &  & (1-\acute{s})
\end{array}
}}{\left( 
\begin{array}{cc}
(K,K) & (L,M) \\ 
(M,L) & (N,N)
\end{array}
\right) }
\end{equation}
In this game Alice plays $S_{1}$ with probability of heads $r$, and Bob
plays $\acute{S}_{2}$ with probability of heads $\acute{s}$. The other
equations in (\ref{bilinear payoffs}) can be given similar interpretation.

We now find constraints on four-coin statistics (\ref{4 Coin Stats}) that
make the payoffs of Eq. (\ref{Payoffs}) identical to the bi-linear payoffs
of Eq. (\ref{bilinear payoffs}) for any real numbers $K,L,M$ and $N$. A
comparison of these equations shows that it happens when $r,s,\acute{r}$ and 
$\acute{s}$ depend on $p_{i}$, for $16\geqslant i\geqslant 0$, as follows

\begin{equation}
\left. 
\begin{array}{c}
r\acute{r}=p_{1},\text{ \ \ }r(1-\acute{r})=p_{2},\text{ \ \ }\acute{r}%
(1-r)=p_{3},\text{ \ \ }(1-r)(1-\acute{r})=p_{4} \\ 
r\acute{s}=p_{5},\text{ \ \ }r(1-\acute{s})=p_{6},\text{ \ \ }\acute{s}%
(1-r)=p_{7},\text{ \ \ }(1-r)(1-\acute{s})=p_{8} \\ 
s\acute{r}=p_{9},\text{ \ \ }s(1-\acute{r})=p_{10},\text{ \ \ }\acute{r}%
(1-s)=p_{11},\text{ \ \ }(1-s)(1-\acute{r})=p_{12} \\ 
s\acute{s}=p_{13},\text{ \ \ }s(1-\acute{s})=p_{14},\text{ \ \ }\acute{s}%
(1-s)=p_{15},\text{ \ \ }(1-s)(1-\acute{s})=p_{16}
\end{array}
\right\}  \label{probability definitions}
\end{equation}
The probabilities $r,s,\acute{r}$ and $\acute{s}$ can be read from (\ref
{probability definitions}) as

\begin{equation}
r=p_{1}+p_{2}\text{, \ \ }s=p_{9}+p_{10}\text{, \ \ }\acute{r}=p_{1}+p_{3}%
\text{, \ \ }\acute{s}=p_{5}+p_{7}  \label{discrete prob}
\end{equation}
provided that $p_{i}$ satisfy

\begin{equation}
\left. 
\begin{array}{c}
p_{1}+p_{2}=p_{5}+p_{6},\text{ \ \ }p_{1}+p_{3}=p_{9}+p_{11} \\ 
p_{9}+p_{10}=p_{13}+p_{14},\text{ \ \ }p_{5}+p_{7}=p_{13}+p_{15}
\end{array}
\right\}  \label{constraints on probs}
\end{equation}
With the defining relations (\ref{discrete prob}), and the constraints (\ref
{constraints on probs}) on the four-coin statistics, each pair in $(S_{1},%
\acute{S}_{1}),(S_{1},\acute{S}_{2}),(S_{2},\acute{S}_{1})$ and $(S_{2},%
\acute{S}_{2})$ gains interpretation of a mixed strategy game. The
correspondence (\ref{correspondences}) means that, for example, Alice's
payoffs read as

\begin{equation}
\left. 
\begin{array}{c}
P_{A}(S_{1},\acute{S}_{1})=P_{A}(r,\acute{r})\text{, \ \ }P_{A}(S_{1},\acute{%
S}_{2})=P_{A}(r,\acute{s}) \\ 
P_{A}(S_{2},\acute{S}_{1})=P_{A}(s,\acute{r})\text{, \ \ }P_{A}(S_{2},\acute{%
S}_{2})=P_{A}(s,\acute{s})
\end{array}
\right\}
\end{equation}
Now, suppose $(s,\acute{s})$ is a Nash equilibrium (NE) i.e.

\begin{equation}
\left\{ P_{A}(s,\acute{s})-P_{A}(r,\acute{s})\right\} \geqslant 0\text{, \ \ 
}\left\{ P_{B}(s,\acute{s})-P_{B}(s,\acute{r})\right\} \geqslant 0
\label{NE}
\end{equation}
Using Eqs. (\ref{bilinear payoffs}) one gets

\begin{equation}
\left. 
\begin{array}{c}
P_{A}(s,\acute{s})-P_{A}(r,\acute{s})=(s-r)\left\{ (K-M-L+N)\acute{s}%
+(L-N)\right\} \geqslant 0 \\ 
P_{B}(s,\acute{s})-P_{B}(s,\acute{r})=(\acute{s}-\acute{r})\left\{
(K-M-L+N)s+(L-N)\right\} \geqslant 0
\end{array}
\right\}  \label{NE1}
\end{equation}
Consider the game of prisoners' dilemma (PD) which is produced when $M>K>N>L$
in the matrix (\ref{matrix}). We select our first representation of PD by
taking

\begin{equation}
K=3,\text{ }L=0,\text{ }M=5\text{ \ and \ }N=1  \label{PDConstants1}
\end{equation}
and the inequalities (\ref{NE1}) are reduced to

\begin{equation}
0\geqslant (s-r)(1+\acute{s}),\text{ \ \ }0\geqslant (\acute{s}-\acute{r}%
)(1+s)  \label{NEPD}
\end{equation}
Now from (\ref{NE1}) the pair $(s,\acute{s})$ is a NE when both inequalities
in (\ref{NEPD}) are true for all $r,\acute{r}\in \lbrack 0,1]$. Because $%
(1+s)\geqslant 1$ and $(1+\acute{s})\geqslant 1$, it produces $s=\acute{s}=0$
as the equilibrium. In present set-up to play PD this equilibrium appears
if, apart from the constraints of Eqs. (\ref{discrete prob}, \ref
{constraints on probs}), we also have

\begin{equation}
s=p_{9}+p_{10}=0,\text{ \ \ }\acute{s}=p_{5}+p_{7}=0
\label{four-coin constraints2}
\end{equation}
which are other constraints on the four-coins statistics (\ref{4 Coin Stats}%
) to hold true if the PD produces the NE $(s,\acute{s})=(0,0)$.

The above analysis can be reproduced for other probability pairs in (\ref
{probability pairs}). For example, when $(r,\acute{s})$ is NE i.e.

\begin{equation}
\left\{ P_{A}(r,\acute{s})-P_{A}(s,\acute{s})\right\} \geqslant 0\text{, \ \ 
}\left\{ P_{B}(r,\acute{s})-P_{B}(r,\acute{r})\right\} \geqslant 0
\label{2ndNE}
\end{equation}
and we again get $(r,\acute{s})=(0,0)$. The relations (\ref{discrete prob})
though now say that it will exist as a NE when

\begin{equation}
p_{1}+p_{2}=0\text{, \ \ }p_{5}+p_{7}=0
\end{equation}
which should, of course, be true along with the relations (\ref{constraints
on probs}). That is, in order to reproduce a particular NE in the bi-matrix
game the probabilities of heads of the four coins representing the game need
to be fixed. Also, from the bi-linear payoffs (\ref{bilinear payoffs}) it is
clear that at the equilibria $(s,\acute{s})=(0,0)$ and $(r,\acute{s})=(0,0)$
the reward for both the players is $N$.

Summarizing, we have shown that when the four-coin statistics (\ref{4 Coin
Stats}) satisfy the constraints of Eqs. (\ref{Constraints}, \ref{constraints
on probs}), the payoffs (\ref{Payoffs}) can be interpreted in terms of a
mixed strategy version of a bi-matrix game. In this setting four strategies $%
S_{1},S_{2},\acute{S}_{1}$ and $\acute{S}_{2}$ available to the players are
associated with the probabilities $r,s,\acute{r}$ and $\acute{s}$,
respectively. This association allows to interpret the payoff recipe of Eq. (%
\ref{Payoffs}) in terms of a mixed strategy game. We showed that when $r,s,%
\acute{r}$ and $\acute{s}$ are expressed in terms of the probabilities $p_{i}
$ for $16\geqslant i\geqslant 1$ (as it is the case in the Eqs. (\ref
{discrete prob})) the bi-linear payoffs (\ref{bilinear payoffs}) become
identical to the payoffs of the Eq. (\ref{Payoffs}). This procedure is
designed to re-express playing a bi-matrix game with four coins in a way
that choosing which coin to toss is a player's strategy.

\section{\label{Games with perfectly correlated}Games with perfectly
correlated particles}

The re-expression of playing a bi-matrix game in terms of a four-coin
tossing experiment, performed between two players, opens the way to see what
happens when the four-coin statistics become correlated. Especially, what if
the correlations go beyond what is achievable with the so-called classical
`coins'.

Presently, there appears general agreement in quantum physics community that
the EPR-type experiments, performed on correlated pairs of particles,
violate the predictions of LHV models. Negative probabilities are found to
emerge when certain LHV models are forced to predict the experimental
outcomes of EPR-type experiments. For example, Han, Hwang and Koh \cite{Han
et al} showed the need for negative probabilities when explicit solutions
can reproduce quantum mechanical predictions for some spin-measurement
directions for all entangled states. In Han et al.'s analysis a special
basis is used to show the appearance of negative probabilities for a class
of LHV models.

Rothman and Sudarshan \cite{Rothman} demonstrated that quantum mechanics
does predict a set of probabilities that violate the CHSH inequality;
however these probabilities are not positive-definite. Nevertheless, they
are physically meaningful in that they give the usual quantum-mechanical
predictions in physical situations. Rothman and Sudarshan observed that all
derivations of Bell's inequalities assume that LHV theories produce a set of
positive-definite probabilities for detecting a particle with a given spin
orientation.

Using a similar approach, Cereceda \cite{Cereceda} proved independently the
necessity of negative probabilities in \textit{all} instances where the
predictions of the LHV model are made to violate the Bell-CHSH inequality.
Interestingly, Cereceda's proof does not rely on any particular basis states
or measurement directions. In the concluding section of his paper Cereceda
analyzes the case of pairs of particles that have perfect correlation
between them. He then proceeds to show the necessity of negative
probabilities for those pairs.

The necessity of negative probability measures, to explain the experimental
outcomes in EPR-type experiments, motivates questions about the effects and
consequences these may have on solution of a game which is physically
implemented using such experiments. This question can be expressed as
follows. What happens to the players' payoffs and solutions of a game that
is physically implemented using pairs of perfectly correlated particles? It
seems quite reasonable to demand that, when the predictions of LHV model
agree with the Bell-CHSH inequality, the game attains classical
interpretation.

One clear advantage of the above approach towards quantum games appears to
be that it is possible to see, without using the machinery of quantum
mechanics, how game-theoretical solutions are affected when a game is
physically implemented using quantum mechanically correlated pairs of
particles.

We follow Cereceda's notation \cite{Cereceda} for probabilities in EPR-type
experiments designed to test Bell-CHSH inequality. Two correlated particles $%
1$ and $2$ are emitted in opposite directions from a common source.
Afterwards, each of the particles enters its own measuring apparatus which
can measure either one of two physical variables at a time. We denote these
variables $S_{1}$ or $S_{2}$ for particle $1$ and $\acute{S}_{1}$ or $\acute{%
S}_{2}$ for particle $2$. These variables can take possible values of $+1$
and $-1$. The source emits very large number of particle pairs.

To describe this experiment Cereceda \cite{Cereceda} considers a
deterministic hidden variable model as follows. The model assumes that there
exists a hidden variable $\lambda $ for every pair of particles emitted by
the source. $\lambda $ has a domain of variation $\Lambda $ and it
determines locally (for example, at the common source) the response of the
particle to each of the measurements they can be subjected to. It is
possible to partition the set of all $\lambda $ into $16$ disjoint subsets $%
\Lambda _{i}$ (with respect to probability measure $m_{i}$) according to the
outcomes of the four possible measurements, $S_{1}$ or $S_{2}$ for particle $%
1$ and $\acute{S}_{1}$ or $\acute{S}_{2}$ for particle $2$. Table $1$ is
reproduced from Cereceda's paper. It shows the $16$ rows characterizing the
subsets $\Lambda _{i}$. The $i$th row gives the outcome of different
measurements when the particle pair is described by a hidden variable
pertaining to the subset $\Lambda _{i}$.

Table $1$: The set $\Lambda $ is partitioned into $16$ possible subsets. The
hidden variables in each subset $\Lambda _{i}$ uniquely determine the
outcomes for each of the four possible single measurements $S_{1},\acute{S}%
_{1},S_{2},$ and $\acute{S}_{2}$. The table is reproduced from the Ref. \cite
{Cereceda}.

\[
\begin{tabular}{lll}
Subset of $\Lambda $ & $
\begin{array}{cccc}
S_{1} & \acute{S}_{1} & S_{2} & \acute{S}_{2}
\end{array}
$ & {\small Probability measure} \\ 
$
\begin{array}{c}
\Lambda _{1} \\ 
\Lambda _{2} \\ 
\Lambda _{3} \\ 
\Lambda _{4} \\ 
\Lambda _{5} \\ 
\Lambda _{6} \\ 
\Lambda _{7} \\ 
\Lambda _{8} \\ 
\Lambda _{9} \\ 
\Lambda _{10} \\ 
\Lambda _{11} \\ 
\Lambda _{12} \\ 
\Lambda _{13} \\ 
\Lambda _{14} \\ 
\Lambda _{15} \\ 
\Lambda _{16}
\end{array}
$ & \multicolumn{1}{c}{$
\begin{array}{cccc}
+ & + & + & + \\ 
+ & + & + & - \\ 
+ & + & - & + \\ 
+ & + & - & - \\ 
+ & - & + & + \\ 
+ & - & + & - \\ 
+ & - & - & + \\ 
+ & - & - & - \\ 
- & + & + & + \\ 
- & + & + & - \\ 
- & + & - & + \\ 
- & + & - & - \\ 
- & - & + & + \\ 
- & - & + & - \\ 
- & - & - & + \\ 
- & - & - & -
\end{array}
$} & $
\begin{array}{c}
m_{1} \\ 
m_{2} \\ 
m_{3} \\ 
m_{4} \\ 
m_{5} \\ 
m_{6} \\ 
m_{7} \\ 
m_{8} \\ 
m_{9} \\ 
m_{10} \\ 
m_{11} \\ 
m_{12} \\ 
m_{13} \\ 
m_{14} \\ 
m_{15} \\ 
m_{16}
\end{array}
$%
\end{tabular}
\]
The probabilities $p_{i}$ are given below in obvious notation \cite{Cereceda}%
.

\begin{eqnarray}
p_{1} &\equiv &p(S_{1}+;\acute{S}_{1}+)=m_{1}+m_{2}+m_{3}+m_{4},  \label{Eq1}
\\
p_{2} &\equiv &p(S_{1}+;\acute{S}_{1}-)=m_{5}+m_{6}+m_{7}+m_{8},  \label{Eq2}
\\
p_{3} &\equiv &p(S_{1}-;\acute{S}_{1}+)=m_{9}+m_{10}+m_{11}+m_{12},
\label{Eq3} \\
p_{4} &\equiv &p(S_{1}-;\acute{S}_{1}-)=m_{13}+m_{14}+m_{15}+m_{16}, \\
p_{5} &\equiv &p(S_{1}+;\acute{S}_{2}+)=m_{1}+m_{3}+m_{5}+m_{7},  \label{Eq5}
\\
p_{6} &\equiv &p(S_{1}+;\acute{S}_{2}-)=m_{2}+m_{4}+m_{6}+m_{8},  \label{Eq6}
\\
p_{7} &\equiv &p(S_{1}-;\acute{S}_{2}+)=m_{9}+m_{11}+m_{13}+m_{15}, \\
p_{8} &\equiv &p(S_{1}-;\acute{S}_{2}-)=m_{10}+m_{12}+m_{14}+m_{16}, \\
p_{9} &\equiv &p(S_{2}+;\acute{S}_{1}+)=m_{1}+m_{2}+m_{9}+m_{10}, \\
p_{10} &\equiv &p(S_{2}+;\acute{S}_{1}-)=m_{5}+m_{6}+m_{13}+m_{14}, \\
p_{11} &\equiv &p(S_{2}-;\acute{S}_{1}+)=m_{3}+m_{4}+m_{11}+m_{12}, \\
p_{12} &\equiv &p(S_{2}-;\acute{S}_{1}-)=m_{7}+m_{8}+m_{15}+m_{16}, \\
p_{13} &\equiv &p(S_{2}+;\acute{S}_{2}+)=m_{1}+m_{5}+m_{9}+m_{13}, \\
p_{14} &\equiv &p(S_{2}+;\acute{S}_{2}-)=m_{2}+m_{6}+m_{10}+m_{14}, \\
p_{15} &\equiv &p(S_{2}-;\acute{S}_{2}+)=m_{3}+m_{7}+m_{11}+m_{15}, \\
p_{16} &\equiv &p(S_{2}-;\acute{S}_{2}-)=m_{4}+m_{8}+m_{12}+m_{16}
\label{Eq16}
\end{eqnarray}
Combining Eqs. (\ref{Constraints}) with Table $1$ gives

\begin{equation}
\stackunder{i=1}{\stackrel{16}{\sum }}m_{i}=1  \label{Sum m's}
\end{equation}
Continuing with Cereceda's description \cite{Cereceda} an example is now
considered. Suppose the particle pair is described by a given $\lambda \in
\Lambda _{2}$, then the particles must behave as follows. If $S_{1}$ is
measured on particle $1$ the result will be $+1$, if $S_{2}$ is measured on
particle $1$ the result will be $+1$, if $\acute{S}_{1}$ is measured on
particle $2$ the result will be $+1$, if $\acute{S}_{2}$ is measured on
particle $2$ the result will be $-1$. Also for each of the plans the results
of measurements made on particle $1$ are independent of the results of
measurements made on particle $2$.

For perfectly correlated particles two of the probabilities $p_{2}$ and $%
p_{3}$ can be set equal to zero. Physically it means that the results for
the joint measurement of two observables, one for each particle, must both
be either $+1$ or $-1$. From a physical point of view, it is reasonable to
suppose that, for the case in which $p_{2}=0$ and $p_{3}=0$, the probability
measures $m_{5},m_{6},m_{7},m_{8},m_{9},m_{10},m_{11},$ and $m_{12}$ also
vanish. It can be verified form Eqs. (\ref{Eq2}, \ref{Eq3}). If this is not
the case then joint detection events will be generated by the LHV model,
which do not happen according to the assumptions made. Cereceda showed that,
in this case, when the predictions of the LHV model violate the Bell-CHSH
inequality, the negativity of either $m_{4}$ or $m_{13}$ can be proved.

We assume now that the probabilities $p_{i}$ appearing in Eqs. (\ref{Payoffs}%
) correspond to the LHV model of the EPR-type experiments performed to test
Bell-CHSH inequality. Like it is the case with the four-coin tossing
experiment, the payoffs (\ref{Payoffs}) can be interpreted as bi-linear
payoffs. To do this we use the Eqs. (\ref{discrete prob}, \ref{constraints
on probs}) to get

\begin{equation}
r=p_{1}=p_{5}+p_{6},\text{ \ \ }\acute{r}=p_{1}=p_{9}+p_{11},\text{ \ \ }%
s=p_{9}+p_{10},\text{ \ \ }\acute{s}=p_{5}+p_{7}
\label{Probabilities Referee}
\end{equation}
But $p_{2}=p_{3}=0$, so that from the first Eq. of (\ref{probability
definitions}) we have $r(1-\acute{r})=\acute{r}(1-r)$ which gives $p_{1}=0$
or $1$ because $r=\acute{r}=p_{1}$ from (\ref{Probabilities Referee}). We
select\footnote{%
Selecting $p_{1}=0$ makes $p_{4}=1$ because $\stackunder{i=1}{\stackrel{4}{%
\sum }}p_{i}=1$. It will result in different but analogous expressions for $%
r,\acute{r},s$ and $\acute{s}$ given in terms of $m_{i}$ without affecting
the present argument.} $p_{1}=1$. Now, from Eqs. (\ref{Eq5}, \ref{Eq6}) we
have $p_{5}+p_{6}=\stackunder{i=1}{\stackrel{8}{\sum }}m_{i}$ and because $%
\stackunder{i=1}{\stackrel{16}{\sum }}m_{i}=1$ from Eq. (\ref{Sum m's}) and $%
m_{i}=0$ for $12\geqslant i\geqslant 5$ we get

\begin{equation}
\left. 
\begin{array}{c}
r=\acute{r}=1=m_{1}+m_{2}+m_{3}+m_{4} \\ 
s=m_{1}+m_{2}+m_{13}+m_{14} \\ 
\acute{s}=m_{1}+m_{3}+m_{13}+m_{15}
\end{array}
\right\}  \label{ProbsCorrelParticles}
\end{equation}
With these relations the bi-linear payoffs (\ref{bilinear payoffs}) are
written as

\begin{eqnarray}
P_{A}(S_{1},\acute{S}_{1}) &=&K  \label{CorrelPayoff1} \\
P_{A}(S_{1},\acute{S}_{2}) &=&L+(K-L)(m_{1}+m_{3}+m_{13}+m_{15})
\label{CorrelPayoff2} \\
P_{A}(S_{2},\acute{S}_{1}) &=&M+(K-M)(m_{1}+m_{2}+m_{13}+m_{14})
\label{CorrelPayoff3}
\end{eqnarray}

\begin{equation}
\left. 
\begin{array}{c}
P_{A}(S_{2},\acute{S}_{2})=(K-L-M+N)\times \\ 
(m_{1}+m_{2}+m_{13}+m_{14})(m_{1}+m_{3}+m_{13}+m_{15})+ \\ 
(L-N)(m_{1}+m_{2}+m_{13}+m_{14})+ \\ 
(M-N)(m_{1}+m_{3}+m_{13}+m_{15})+N
\end{array}
\right\}  \label{CorrelPayoff4}
\end{equation}
Each of the correlated payoffs (\ref{CorrelPayoff2}, \ref{CorrelPayoff3}, 
\ref{CorrelPayoff4}) can be split into two parts i.e.

\begin{equation}
\left. 
\begin{array}{c}
P_{A}(S_{1},\acute{S}_{2})=P_{A_{a}}(S_{1},\acute{S}_{2})+P_{A_{b}}(S_{1},%
\acute{S}_{2}) \\ 
P_{A}(S_{2},\acute{S}_{1})=P_{A_{a}}(S_{2},\acute{S}_{1})+P_{A_{b}}(S_{2},%
\acute{S}_{1}) \\ 
P_{A}(S_{2},\acute{S}_{2})=P_{A_{a}}(S_{2},\acute{S}_{2})+P_{A_{b}}(S_{2},%
\acute{S}_{2})
\end{array}
\right\}  \label{payoff splitting}
\end{equation}
where

\begin{eqnarray}
P_{A_{a}}(S_{1},\acute{S}_{2}) &=&L+(K-L)(m_{1}+m_{3})
\label{Alice's split payoffs1} \\
P_{A_{b}}(S_{1},\acute{S}_{2}) &=&(K-L)(m_{13}+m_{15})
\label{Alice's split payoffs2} \\
P_{A_{a}}(S_{2},\acute{S}_{1}) &=&M+(K-M)(m_{1}+m_{2})
\label{Alice's split payoffs3} \\
P_{A_{b}}(S_{2},\acute{S}_{1}) &=&(K-M)(m_{13}+m_{14})
\label{Alice's split payoffs41}
\end{eqnarray}

\begin{equation}
\left. 
\begin{array}{c}
P_{A_{a}}(S_{2},\acute{S}_{2})=(K-L-M+N)(m_{1}+m_{2})(m_{1}+m_{3})+ \\ 
(L-N)(m_{1}+m_{2})+(M-N)(m_{1}+m_{3})+N
\end{array}
\right\}  \label{Alice's split payoffs5}
\end{equation}

\begin{equation}
\left. 
\begin{array}{c}
P_{A_{b}}(S_{2},\acute{S}_{2})=(K-L-M+N)\{(m_{1}+m_{2})(m_{13}+m_{15})+ \\ 
(m_{1}+m_{3})(m_{13}+m_{14})+(m_{13}+m_{14})(m_{13}+m_{15})\}+ \\ 
(L-N)(m_{13}+m_{14})+(M-N)(m_{13}+m_{15})
\end{array}
\right\}  \label{Alice's split payoffs6}
\end{equation}
The significance of this splitting is that components $P_{A_{b}}(S_{1},%
\acute{S}_{2}),$ $P_{A_{b}}(S_{2},\acute{S}_{1})$ and $P_{A_{b}}(S_{2},%
\acute{S}_{2})$ of the Alice's payoffs in Eqs. (\ref{payoff splitting})
become zero when the predictions of LHV model agree with the Bell-CHSH
inequality. Cheon and Tsutsui \cite{Cheon} have also shown a similar
splitting using correlated payoff operators whose expectation values are the
players' payoffs.

Consider again the PD with the selection of the constants given in (\ref
{PDConstants1}). Let the game be played using correlated particles, for
which substitutions can be made from (\ref{ProbsCorrelParticles}) into the
Nash equilibrium condition (\ref{NEPD}). It gives

\begin{equation}
0\geqslant (s-1)(1+\acute{s}),\text{ \ \ }0\geqslant (\acute{s}-1)(1+s)
\label{NEPD1}
\end{equation}
where

\begin{eqnarray}
s &=&m_{1}+m_{2}+m_{13}+m_{14}  \label{def of s} \\
\acute{s} &=&m_{1}+m_{3}+m_{13}+m_{15}  \label{def of s'}
\end{eqnarray}
It can be noticed that

\begin{itemize}
\item  The predictions of the LHV model agree with the Bell-CHSH inequality.
It makes $m_{i}\geqslant 0$ for all $16\geqslant i\geqslant 1$. Combining it
with (\ref{ProbsCorrelParticles}), i.e.$\stackunder{i=1}{\stackrel{4}{\text{ 
}\sum m_{i}}}=1,$ gives

\begin{equation}
m_{13}=m_{14}=m_{15}=m_{16}=0  \label{m13 to m16=0}
\end{equation}
\end{itemize}

So that, it reduces $s$ and $\acute{s}$ in (\ref{def of s}, \ref{def of s'})
to

\begin{equation}
s_{1}=m_{1}+m_{2},\text{ \ \ }\acute{s}_{1}=m_{1}+m_{3}  \label{New Def s s'}
\end{equation}

\begin{itemize}
\item  The requirement $(s_{1},\acute{s}_{1})=(0,0)$ says that when the
predictions of the LHV model agree with the Bell-CHSH inequality the pair $%
(0,0)$ is a NE. Eq. (\ref{New Def s s'}) gives
\end{itemize}

\begin{equation}
m_{1}=m_{2}=m_{3}=0  \label{00 is NE when LHV agrees chsh}
\end{equation}
Before proceeding to raise a question, we make following observations:

\begin{enumerate}
\item  Eqs. (\ref{Eq1}) to (\ref{Eq16}) give the probabilities $p_{i}$ in
terms of $m_{i}$, for $16\geqslant i\geqslant 1$, corresponding to the
EPR-type experiments. These probabilities satisfy the constraints (\ref
{constraints on probs}) that emerge when Eqs. (\ref{Payoffs}) are
interpreted in terms of bi-linear payoffs of Eqs. (\ref{bilinear payoffs}).

\item  The expressions (\ref{ProbsCorrelParticles}) for $r,s,\acute{r}$ and $%
\acute{s}$ are obtained from the corresponding expressions for the coin
tossing case (\ref{discrete prob}), while taking into consideration the
constraints on probabilities for perfectly correlated particles.

\item  The Eqs. (\ref{m13 to m16=0}, \ref{00 is NE when LHV agrees chsh})
together make $s=\acute{s}=0$ in the definitions (\ref{def of s}, \ref{def
of s'}). These definitions correspond to the representation (\ref
{PDConstants1}) of PD, for which constraints (\ref{four-coin constraints2})
should be true in case $s=\acute{s}=0$ is a NE when the game is played with
coins. It is observed that both $(p_{9}+p_{10})$ and $(p_{5}+p_{7})$ become
zero from the definitions of $p_{i}$ in terms of $m_{i}$ given in Table $1$,
when Eqs. (\ref{m13 to m16=0}, \ref{00 is NE when LHV agrees chsh}) both are
true. It means that when PD gives $s=\acute{s}=0$ as an equilibrium the
constraints on probabilities become identical in the following two cases:

a) The game is played using repeated tosses with four coins.

b) The game is played with perfectly correlated particles such that the
predictions of LHV model agree with the Bell-CHSH inequality.
\end{enumerate}

Bi-matrix games other than PD, presumably with different Nash equilibria,
would give rise to different but analogous constraints on the probabilities $%
s,\acute{s},r$ and $\acute{r}$. In the light of these observations following
question arises immediately. What happens to the Nash conditions (\ref{NEPD1}%
) when the predictions of LHV model disagree with the Bell-CHSH inequality?
To answer it consider the Nash conditions (\ref{NEPD1}) with a substitution
from (\ref{00 is NE when LHV agrees chsh}). These give

\begin{equation}
\left. 
\begin{array}{c}
0\geqslant (s_{2}-1)(1+\acute{s}_{2}) \\ 
0\geqslant (\acute{s}_{2}-1)(1+s_{2})
\end{array}
\right\}  \label{NEPD2}
\end{equation}
where

\begin{equation}
s_{2}=m_{13}+m_{14},\text{ \ \ }\acute{s}_{2}=m_{13}+m_{15}
\label{NE when LHV violates chsh}
\end{equation}
Now we recall that in Cereceda's analysis $m_{13}$ can take negative value
when the predictions of LHV model disagree with the Bell-CHSH inequality.
Both $(s_{2}-1)$ and $(\acute{s}_{2}-1)$ in (\ref{NEPD2}) remain negative
whether $m_{13}$ is positive or negative. Similarly, both $(1+\acute{s}_{2})$
and $(1+s_{2})$ remain positive whether $m_{13}$ is positive or negative.
Therefore, the Nash conditions (\ref{NEPD2}), that corresponds to the
representation (\ref{PDConstants1}) of PD, are \emph{not} violated whether
the predictions of the LHV model agree or disagree with the Bell-CHSH
inequality.

Players' payoffs at the equilibrium $(s_{2},\acute{s}_{2})$ can be found
from Eq. (\ref{Alice's split payoffs6}) as

\begin{equation}
\left. 
\begin{array}{c}
P_{A_{a}}(S_{2},\acute{S}_{2})=P_{B_{a}}(S_{2},\acute{S}_{2})=N \\ 
P_{A_{b}}(S_{2},\acute{S}_{2})=(K-L-M+N)s_{2}\acute{s}_{2}+(L-N)s_{2}+(M-N)%
\acute{s}_{2} \\ 
P_{B_{b}}(S_{2},\acute{S}_{2})=(K-L-M+N)s_{2}\acute{s}_{2}+(M-N)s_{2}+(L-N)%
\acute{s}_{2}
\end{array}
\right\}
\end{equation}
where $s_{2}$ and $\acute{s}_{2}$ are read from (\ref{NE when LHV violates
chsh}). For example, with PD's representation (\ref{PDConstants1}), the
players' payoffs are obtained from Eqs. (\ref{Alice's split payoffs5},\ref
{Alice's split payoffs6}) as

\begin{equation}
\left. 
\begin{array}{c}
P_{A}(S_{2},\acute{S}_{2})=\acute{s}_{2}(4-s_{2})-s_{2}+1 \\ 
P_{B}(S_{2},\acute{S}_{2})=s_{2}(4-\acute{s}_{2})-\acute{s}_{2}+1
\end{array}
\right\}  \label{PD(1) payoffs (Q)}
\end{equation}

The NE $(s_{2},\acute{s}_{2})$ in Eqs. (\ref{NE when LHV violates chsh})
corresponds when the predictions of LHV model disagree with the Bell-CHSH
inequality. Its defining inequalities (\ref{NEPD2}) show that it exists even
when either $s_{2}$ or $\acute{s}_{2}$ take negative values, which can be
realized when $m_{13}$ is negative. So that, the NE in PD's representation (%
\ref{PDConstants1}) can be `displaced' when the predictions of LHV model
disagree with the Bell-CHSH inequality. Here displacement means that either $%
s_{2}$ or $\acute{s}_{2}$ can take negative values. However, this extra
freedom of assuming negative values does \emph{not} disqualify $(s_{2},%
\acute{s}_{2})$ to exist as a NE. From Eq. (\ref{PD(1) payoffs (Q)}) it can
be noticed that $P_{A}(S_{2},\acute{S}_{2})$ and $P_{B}(S_{2},\acute{S}_{2})$
can not be greater than $1$ when both $s_{2}$ and $\acute{s}_{2}$ take
negative values.

We show now that it may not be the case with another representation of PD.
That is, the extra freedom for $s_{2}$ and $\acute{s}_{2}$ to take negative
values, granted when the predictions of the LHV model disagree with
Bell-CHSH inequality, leads to disqualification of $(s_{2},\acute{s}_{2})$
to exist as a NE in that representation of PD.

Consider the PD with a slightly different value assigned to the constant $N$
of the game \cite{Cheon}

\begin{equation}
K=3,\text{ \ }L=0,\text{ \ }M=5\text{ \ and \ }N=0.2  \label{PDConstants2}
\end{equation}
In this representation the inequalities (\ref{NE1}) are reduced to

\begin{equation}
0\geqslant (s-r)(1.8\acute{s}+0.2),\text{ \ \ }0\geqslant (\acute{s}-\acute{r%
})(1.8s+0.2)
\end{equation}
A substitution of $r=\acute{r}=1$ from (\ref{ProbsCorrelParticles}) and then
addition of both the inequalities gives

\begin{equation}
\frac{1}{9}\left\{ 4(s+\acute{s})+1\right\} \geqslant s\acute{s}
\label{CheonGameNE}
\end{equation}
Suppose that the predictions of LHV model disagree the Bell-CHSH inequality
i.e. both $s$ and $\acute{s}$ are to be replaced by $s_{2}$ and $\acute{s}%
_{2}$ in (\ref{CheonGameNE})

\begin{equation}
\frac{1}{9}\left\{ 4(s_{2}+\acute{s}_{2})+1\right\} \geqslant s_{2}\acute{s}%
_{2}  \label{CheonGameNE1}
\end{equation}
where $s_{2}$ and $\acute{s}_{2}$ are given by (\ref{NE when LHV violates
chsh}). Interestingly, it is observed that the inequality (\ref{CheonGameNE1}%
) is violated if $-0.25>(s_{2}+\acute{s}_{2})$ and the NE of Eq. (\ref{NE
when LHV violates chsh}) ceases to exist. Of course, it applies to the
representation of PD given by (\ref{PDConstants2}). Players' payoffs are
given as

\begin{equation}
\left. 
\begin{array}{c}
P_{A}(S_{2},\acute{S}_{2})=\acute{s}_{2}(4.8-1.8s_{2})+0.2(1-s_{2}) \\ 
P_{B}(S_{2},\acute{S}_{2})=s_{2}(4.8-1.8\acute{s}_{2})+0.2(1-\acute{s}_{2})
\end{array}
\right\}  \label{PD(2) payoffs (Q)}
\end{equation}
As it is the case with the first representation of PD, the payoffs $%
P_{A}(S_{2},\acute{S}_{2}),P_{B}(S_{2},\acute{S}_{2})$ can not be greater
than $0.2$ when both $s_{2}$ and $\acute{s}_{2}$ taken negative values.

It shows that in the physically implementation of PD, using perfectly
correlated particles, the two representations (\ref{PDConstants1},\ref
{PDConstants2}) behave differently from each other. In representation (\ref
{PDConstants1}) the disagreement of the predictions of LHV model with the
Bell-CHSH inequality leads to a displacement of the NE $(s,\acute{s})$ such
that $s$ and $\acute{s}$ can assume negative values. Displacement occurs but 
$(s,\acute{s})$ continues to exist as a NE.

On the other hand, in the representation (\ref{PDConstants2}) the
disagreement of the predictions of LHV model with the Bell-CHSH inequality
leads to the disappearance of the NE $(s,\acute{s})$ when both $s$ and $%
\acute{s}$ assume negative values and their sum becomes less than $-0.25$.

Alert reader may come up with a `minimalist' interpretation of the present
approach as follows. Constraints (\ref{constraints on probs},\ref{four-coin
constraints2}) are required on four-coin statistics (\ref{4 Coin Stats}) to
make $(s,\acute{s})=(0,0)$ a NE when the PD is played in representation (\ref
{PDConstants1}) with repeated tosses of four coins. When the \emph{same}
game is played with pairs of perfectly correlated particles and the
predictions of LHV model disagree with the Bell-CHSH inequality, we can have 
$(p_{9}+p_{10})$ or $(p_{5}+p_{7})$ becoming negative which denies (\ref
{four-coin constraints2}). If it affects the solution of the game then one
can say that it is because of the change in the underlying probabilities of
our physical system. In our point of view it is \emph{not }the question of
changing the underlying probabilities. On the other hand, it is a procedure
that addresses the question asking what is the true quantum content of
quantum game in the following two steps:

\begin{enumerate}
\item  For perfectly correlated particles, developing an association that
guarantees a classical game results when the predictions of LHV model do not
violate the Bell-CHSH inequality.

\item  With above association retained, how solutions of a game are affected
when the LHV model violates the Bell-CHSH inequality.
\end{enumerate}

When these steps taken into consideration the possibility of construction of
a classical game, that reproduces a quantum game, can not be taken as a blow
to the subject of quantum games \cite{Enk}. In our opinion the question
quantum game theory asks is how quantum mechanical aspects of a physical
system leave their mark on game-theoretic solutions. The possibility of
classical construction of a quantum game does not make the question
disappear.

\section{Concluding remarks}

The results in this paper can be summarized as follows. To establish a
better comparison with EPR-type experiments, a hypothetical physical
implementation of a bi-matrix game is developed that uses repeated tosses
with four coins . This opens the way to introduce directly the peculiar
probabilities involved in the EPR-type experiments, designed to test the
Bell-CHSH inequality, into the proposed procedure to play a bi-matrix game.

The argument rests on the result that when perfect correlations exist
between two particles that are forwarded to two players, the violation of
Bell-CHSH inequality by the predictions of a class of local hidden variable
models forces certain probability measures to take negative values. We
investigate how this aspect affects a game and its solutions when it is
physically implemented using EPR-type set-up. In such a set-up two choices
are made available to each player that are taken as their strategies.
Players' payoffs depend on the outcomes of repeated measurements and the
constants that define the game.

We find that a consistent set of probabilities can be obtained, given in
terms of the statistics involved in the four-coin tossing experiment, such
that the game between the players is interpretable as a classical bi-matrix
game allowing mixed strategies. The proposal is designed in a way that
allows, in the next step, to introduce directly the peculiar probabilities
emerging in the EPR-type experiments.

We find that when the game is played with perfectly correlated pairs of
particles, the players' payoffs are observed to split into two parts; which
correspond to the two situations arising when the predictions, of the class
of local hidden variable models, do and do not violate the Bell-CHSH
inequality, respectively.

Apart from the splitting of the payoffs, we showed that the implementation
using perfectly correlated particles distinguishes between two
representations of the game that are completely equivalent in the classical
context. We observed that the effects on a game-theoretic solution concept,
of whether the predictions of the local hidden variable model do or do not
violate the Bell-CHSH inequality, is sensitive to the particular
representation used for the game.

\bigskip 

{\Large Acknowledgment:}{\large \ }Author is thankful to anonymous referees
for giving helpful comments.

\end{document}